\documentclass[twocolumn,showpacs,10pt]{revtex4-1}

\usepackage{graphicx} 
\usepackage{color}

\usepackage[english]{babel}
\usepackage[utf8]{inputenc}

\usepackage{amssymb}
\usepackage{amsmath}
\usepackage{xspace}
\usepackage{ulem}

\begin{document}

\title[Short Title]{Spin pairs in a weakly coupled many-electron quantum dot}
\author{S.~Hellm\"uller}
\email{hesarah@phys.ethz.ch}
\author{D.~Bischoff}
\author{T.~M\"uller}
\author{M.~Beck}
\author{K.~Ensslin}
\author{T.~Ihn}
\affiliation{Solid State Physics Laboratory, ETH Zurich, 8093 Zurich, Switzerland}
\date{\today}


\begin{abstract}
We report the observation of an unusually large number of consecutive spin pairs in a weakly coupled many-electron $\mathrm{GaAs/AlGaAs}$ quantum dot. The pairs are identified due to pairwise parallel shifts of Coulomb resonances in a perpendicular magnetic field. Using a nearby quantum point contact for time-resolved charge detection, the tunneling rates are investigated as a function of gate voltage and magnetic field. We compare our experimental data to a single-level transport model and discuss possible reasons for deviations.
\end{abstract}


\pacs{72.25.Dc, 73.21.La, 73.23.-b, 73.23.Hk}

\maketitle


\section{Introduction}
Quantum dots (QDs) can be regarded as artificial atoms \cite{kastner1992, kastner1993, reimann2002} with an addition spectrum that depends on the single-particle state defined by the confinement \cite{fock1928, darwin1931} as well as on the Coulomb interaction and the exchange interaction between the electrons \cite{kouwenhoven2001}. A successive filling of a spin up and a spin down electron into the same orbital state is called a spin pair. The addition spectra of quantum dots have been investigated in various materials and for different geometries \cite{tarucha1996, tarucha1997, kouwenhoven1997b, tarucha1998, tans1998, ciorga2000, folk2001, luescher2001, luescher2001b, lindemann2002, buitelaar2002, cobden2002, herrero2004, moriyama2005, guettinger2010}. While for QDs formed in carbon nano\-tubes normally a successive filling with spin pairs is observed \cite{buitelaar2002, cobden2002}, this is usually not the case for QDs in $\mathrm{GaAs/AlGaAs}$ heterostructures \cite{tarucha1996, ciorga2000, folk2001, luescher2001, luescher2001b, lindemann2002}. 
Indeed, the exchange interaction in $\mathrm{GaAs/AlGaAs}$ heterostructures is typically large enough to prevent a successive filling of spin up and spin down electrons into a QD by favoring parallel spins, and hence, spin pairs are observed only occasionally \cite{ciorga2000, lindemann2002}. Signatures of spin-pairing \cite{luescher2001b} were also observed in a statistical analysis \cite{beenakker1997, alhassid2000, alhassid2000b, aleiner2002} of even and odd nearest neighbor peak spacings \cite{folk1996, patel1998}.

In this paper we present the measurement of many consecutive spin-pair candidates in a lateral $\mathrm{GaAs/AlGaAs}$ quantum dot similar to the QDs investigated in previous studies \cite{folk1996, ciorga2000, folk2001, luescher2001, luescher2001b, lindemann2002}. Unexpectedly, pairwise parallel shifts in magnetic field of 20 and more consecutive Coulomb peak pairs \cite{comment1} are observed. Tunneling rates are recorded as a function of magnetic field and detuning of the chemical potential of the QD using time-resolved charge detection techniques.

\xspace
\begin{figure*}[tbp]
        \centering
        \parbox[b]{15cm}{ \includegraphics[width=15cm]{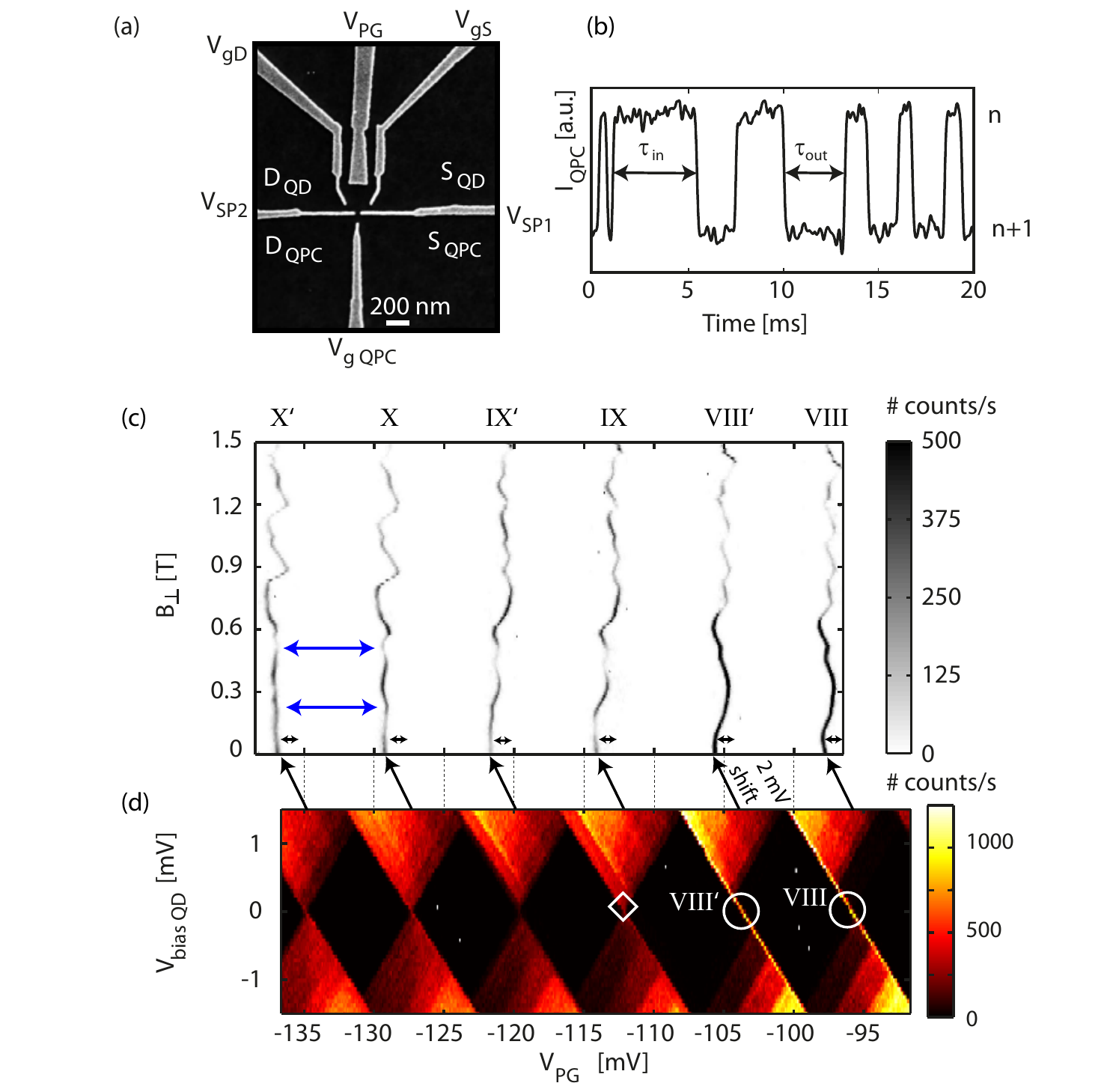}}
        \parbox{15cm}{ \caption{\small (Color online) (a) SEM image of the measured \ensuremath{\mathrm {GaAs/AlGaAs}}\xspace device including a QD and a nearby QPC charge detector. A part of a time-trace ($5\ensuremath{\,\mathrm{kHz}}\xspace$, 8th order Bessel filter) is exemplarily shown in (b) corresponding to a point inside the white diamond in (d). (c) Evolution of some of the Coulomb peaks (number VIII - X') in a perpendicular magnetic field $B$ at zero \ensuremath{\,V_{\mathrm{bias}\,\mathrm{QD}}}\xspace. Plotted is the number of electrons entering and leaving the QD per second as a function of the plunger-gate voltage $V_\mathrm{PG}$. The parallel shift of pairs of neighboring peaks suggests that spin up and spin down electrons are filled pairwise into the QD in this regime. (d) Coulomb diamonds corresponding to the peaks in (c). Plotted is the number of electrons passing the QD per second extracted from $200\,\mathrm{ms}$ long time traces similar to the one shown in (b).
        }
        \label{figure1}}
\end{figure*}

\begin{figure*}[t]
        \centering
        \parbox[b]{15cm}{ \includegraphics[width=15cm]{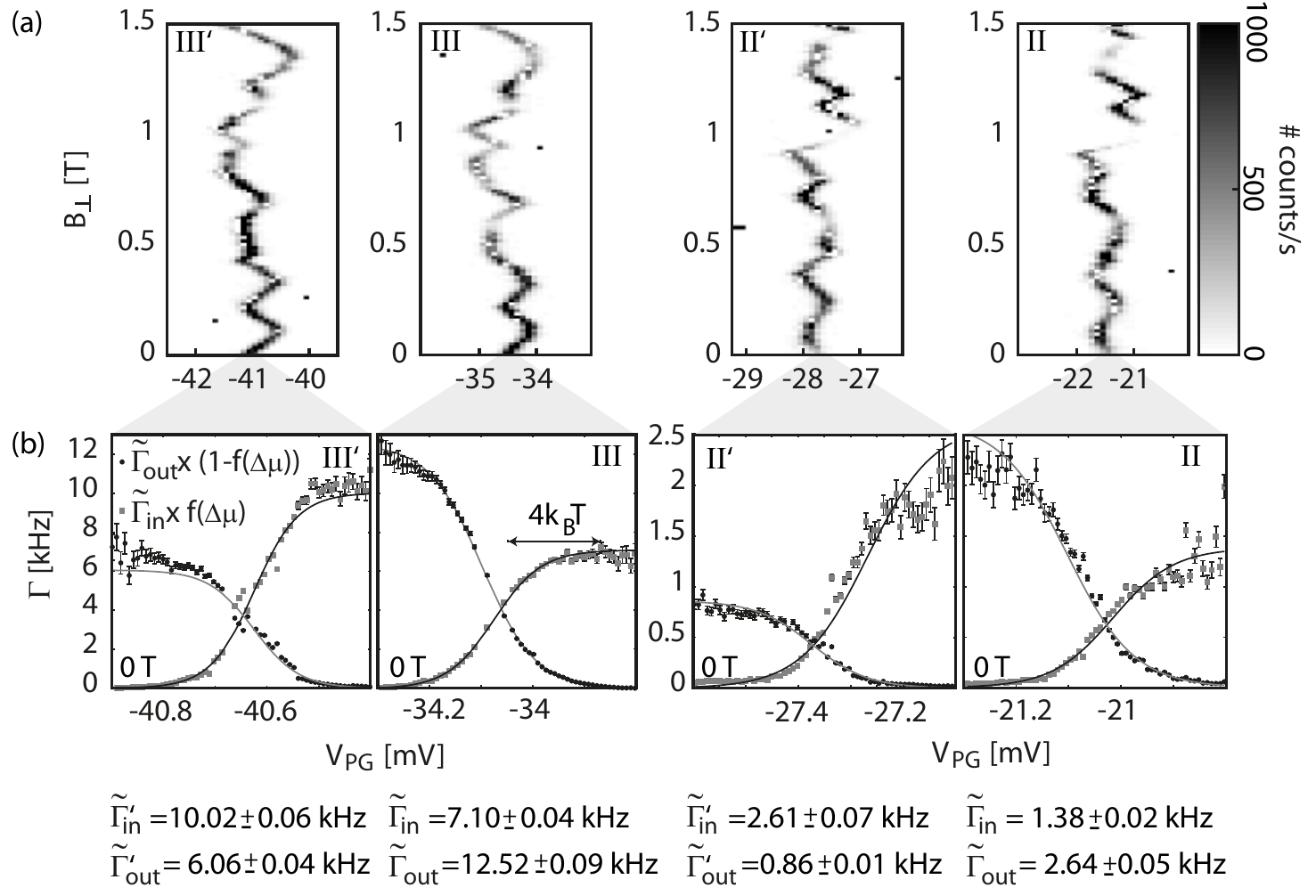}}
        \parbox{15cm}{ \caption{\small (Color online) (a) Evolution of the Coulomb peaks II, II', III and III' in a perpendicular magnetic field $B$ at zero bias. (b) Tunneling rates for the same resonances at zero magnetic field and zero bias. The error bars of the data points correspond to the statistical error $\Gamma/\sqrt{N}$, where $N$ is the number of events. The data is fitted by the single-level expressions $\ensuremath{\Gamma_\mathrm{in}=\tilde{\Gamma}_\mathrm{in}f(\mu)}$ and $\ensuremath{\Gamma_\mathrm{out}=\tilde{\Gamma}_\mathrm{out}(1-f(\mu))}$ using a maximum likelihood method. }
        \label{figure2}}
\end{figure*}

\begin{figure*}[tbp]
        \centering
        \parbox[b]{15cm}{ \includegraphics[width=15cm]{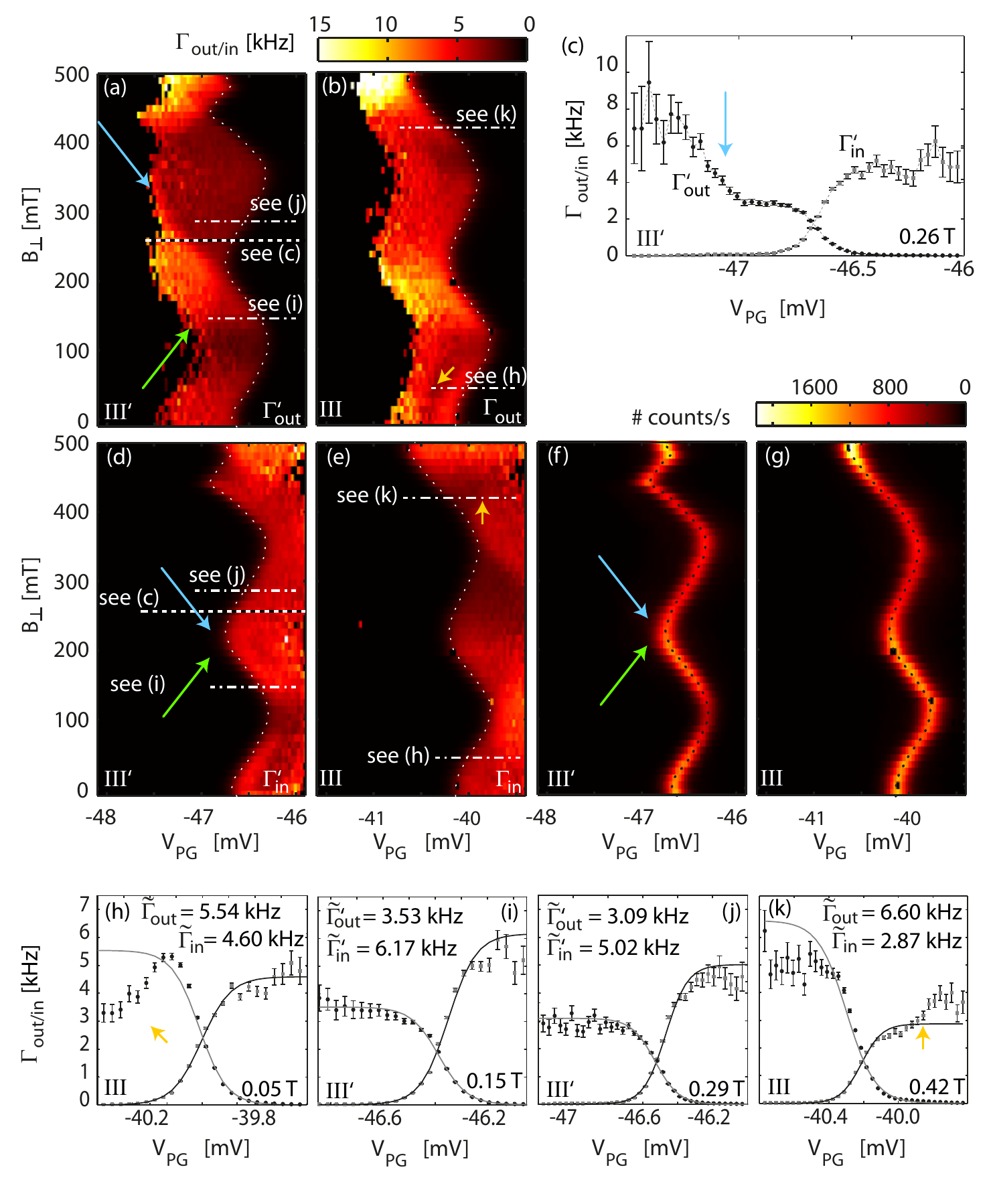}}
        \parbox{15cm}{ \caption{\small (Color online) (a)-(e) The tunneling rates of the Coulomb peak pair III'/III (shown in (f) and (g)). In (a) and (b) the tunneling-out rate, and in (d) and (e) the tunneling-in rate are presented as a function of plunger gate voltage and perpendicular magnetic field $B$ at zero bias voltage ($V_{\mathrm{bias}\,\mathrm{QD}}=0$). (c) A cut along the dotted lines in (a) and (d). The number of energy levels participating in transport decreases stepwise from left to right. The blue and the green arrows mark transitions from one to two energy levels contributing to transport. The energy level marked by blue arrows decreases in energy for increasing magnetic field. The opposite is true for the one marked by a green arrow. The curved, finely dotted lines in (a),(b),(d) and (e) are copies of the dotted black lines in (f) and (g) and mark the position of the Coulomb resonances. (h-k) $\Gamma_\mathrm{out}$ and $\Gamma_\mathrm{in}$ as a function of plunger gate voltage at magnetic fields of (h) $B=0.05\,\mathrm{T}$ (peak III), (i) $B=0.15\,\mathrm{T}$ (peak III'), (j) $B=0.29\,\mathrm{T}$ (peak III') and (k) $B=0.42\,\mathrm{T}$ (peak III). With tunneling rates of (h) \ensuremath{\,\tilde{\Gamma}\xspace _{\mathrm{in}}=4.60\pm0.06\,\mathrm{kHz}}, \ensuremath{\,\tilde{\Gamma}\xspace _{\mathrm{out}}}\xspace= $5.54\pm0.07\,\mathrm{kHz}$, (i) \ensuremath{\,\tilde{\Gamma}\xspace '_{\mathrm{in}}=6.17\pm0.10\,\mathrm{kHz}}, \ensuremath{\,\tilde{\Gamma}\xspace '_{\mathrm{out}}}\xspace= $3.53\pm0.05\,\mathrm{kHz}$, (j) \ensuremath{\,\tilde{\Gamma}\xspace '_{\mathrm{in}}=5.02\pm0.08\,\mathrm{kHz}}, \ensuremath{\,\tilde{\Gamma}\xspace '_{\mathrm{out}}}\xspace= $3.09\pm0.04\,\mathrm{kHz}$ and (k) \ensuremath{\,\tilde{\Gamma}\xspace _{\mathrm{in}}=2.87\pm0.03\,\mathrm{kHz}}, \ensuremath{\,\tilde{\Gamma}\xspace _{\mathrm{out}}}\xspace= $6.6\pm0.13\,\mathrm{kHz}$.} 
        \label{figure3}}
\end{figure*}

\begin{figure}[tbp]
        \centering
        \parbox[b]{7.5cm}{ \includegraphics[width=7.5cm]{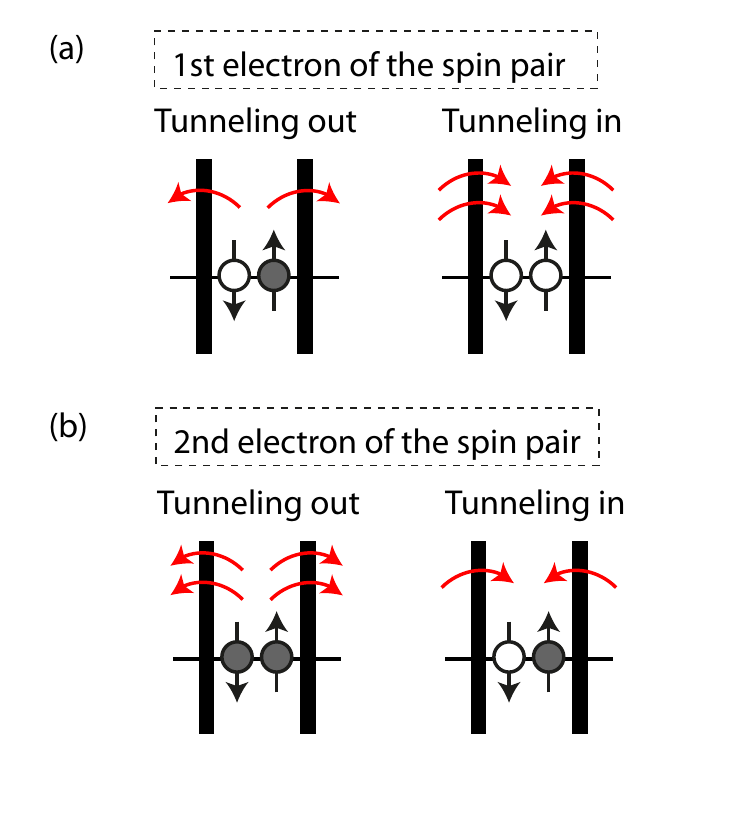}}
        \parbox{7.5cm}{ \caption{\small (Color online) (a,b) Schematics of a model used to explain spin-dependent tunneling rates. The gray (empty) circles indicate filled (empty) states and the red arrows visualize the tunneling rates. The model in (a) describes the tunneling rates of the first electron of the spin pair and the model in (b) describes the tunneling rate of the second electron of the spin pair.}
        \label{figure4}}
\end{figure}

\section{Device and setup}
Our device, shown in Fig.\,\ref{figure1}(a), contains a quantum dot with a nearby quantum point contact (QPC) formed with metal electrodes. They are fabricated by electron beam lithography on a \ensuremath{\mathrm {GaAs/AlGaAs}}\xspace heterostructure with a two-dimensional electron gas (2DEG) located 34\ensuremath{\,\mathrm{nm}}\xspace below the surface. The 2DEG has a mobility of $33\ensuremath{\,\mathrm{m}^2/\mathrm{V}\mathrm{s}}\xspace$ and an electron density of $4.8\times10^{15}\,\ensuremath{\,\mathrm{m}}\xspace^{-2}$ at a temperature of $4\ensuremath{\,\mathrm{K}}\xspace$. The device is measured in a dilution refrigerator at an electron temperature of $T_\mathrm{e}\approx 90\ensuremath{\,\mathrm{mK}}\xspace$ which corresponds to $7.7\,\mu \mathrm{eV}$.

The QD is investigated in a regime where it is populated by at least 160-200 electrons and where the current through the dot is equal to or less than $0.25\ensuremath{\,\mathrm{fA}}\xspace$. Accurate measurements of such small currents are challenging. We therefore use a QPC charge detector capacitively coupled to the QD in order to detect the tunneling of single electrons in a time-resolved manner. A bias voltage  
of $200\ensuremath{\,\mu\mathrm{V}}\xspace$ is applied to the QPC. The resulting time-resolved current is recorded with a sampling rate in the range of $500\,\mathrm{kHz}$ to $1\,\mathrm{MHz}$.
The time-traces contain a random telegraph signal where the steps in the signal correspond to single electrons leaving or entering the QD. 

For the data evaluation it is crucial that the signal-to-noise ratio (SNR) is large enough to avoid false counts. Therefore, after sampling, the data is digitally filtered ($5\ensuremath{\,\mathrm{kHz}}\xspace$ or $15\ensuremath{\,\mathrm{kHz}}\xspace$, 8th order Bessel filter) and resampled before extracting the number of electrons which entered and left the QD during a certain time. From the average times the QD is occupied (unoccupied) by an excess electron, the rate for tunneling out (in) can be determined accurately. A part of a time-trace is exemplarily shown in Fig.\,\ref{figure1}(b). It corresponds to a point inside the white diamond in Fig.\,\ref{figure1}(d).\\

\section{Experimental results}

\subsection{Observation of parallel shift of neighboring Coulomb peaks at finite magnetic fields}
The evolution of the Coulomb peaks in a magnetic field $B$ of up to $1.5\ensuremath{\,\mathrm{T}}\xspace$ is shown in Fig.\,\ref{figure1}(c) for $\ensuremath{\,V_{\mathrm{bias}\,\mathrm{QD}}}\xspace = 0\ensuremath{\,\mathrm{V}}\xspace$. Their positions fluctuate in magnetic fields up to $1.5\ensuremath{\,\mathrm{T}}\xspace$ by $\Delta V_\mathrm{PG}=1.5-3\ensuremath{\,\mathrm{mV}}\xspace$ which is about 15-30\% of their spacing \cite{comment3}. A pairwise correlation is clearly visible, suggesting the occurrence of spin pairs. 
In our sample the perpendicular magnetic field dependence is determined by the orbital wave functions rather than by the spin. Therefore, two parallel shifting peaks indicate the same orbital wave function and therefore opposite spins due to the Pauli principle\cite{comment4}.
Differences in the fluctuations between the first and the second peak of a spin pair\cite{tarucha2000, fuhrer2003} are observed occasionally but are not in the focus of this paper. Two examples of such deviations are visible for peak X and X' at $\approx 0.57\,\mathrm{T}$ and at $\approx 0.27\,\mathrm{T}$ marked by blue arrows in Fig.\,\ref{figure1}(c). 

In total, 46 consecutive Coulomb resonances are investigated and the pairwise correlation for all of them is as clear as for those shown in Fig.\,\ref{figure1}(c). In five cases, however, triples are observed instead of pairs. The triples are most likely pairs where one Coulomb resonance is recorded twice due to corresponding charge rearrangements \cite{comment2}. 
These charge rearrangements can be clearly identified in Coulomb-blockade diamond measurements. They are caused by single charge traps close to the quantum dot and occur at a certain plunger gate voltage.  Additionally, several times over the total measurement period of about 4 months, spontaneous small charge rearrangements happened. For example between the measurement in Fig.\,\ref{figure1}(c) and Fig.\,\ref{figure1}(d), a shift of $2\ensuremath{\,\mathrm{mV}}\xspace$ occurred. The fluctuations and the order of the peaks did not change due to the rearrangements. For small rearrangements no triples where observed  and the positions of all peaks shifted only slightly. However, for large rearrangements, happening at a certain plunger gate voltage, triples were measured and all peaks occurring at larger plunger gate voltages are shifted with respect to the others. \\

\subsection{Coulomb diamonds corresponding to the Coulomb peak pairs}
In Fig.\,\ref{figure1}(d), the number of electrons entering and leaving the QD per second is plotted as a function of the bias voltage applied to the QD and the plunger gate voltage \ensuremath{\,V_\mathrm{pg}}\xspace. In the black areas in the middle of the Coulomb blockade diamonds the number of electrons in the dot is constant. The charging energy is about $1.4\ensuremath{\,\mathrm{meV}}\xspace$ and the single-level spacing is estimated to be $\Delta\mathrm{E}=2\pi\hbar^2/m^*A_\mathrm{dot}\approx 180\ensuremath{\,\mu\mathrm{eV}}\xspace$ with $A_\mathrm{dot}=200\times200\,\mathrm{nm}^2$, which is in agreement with the excited states observed in the measurement. According to Ref.~\onlinecite{lindemann2002}, the ratio of electron-electron interaction energy and Fermi energy, characterized by the electron gas interaction parameter $r_s$, is a measure for the probability to observe spin pairs. If $r_s < 1$, spin pairs are expected to be the norm. However, there is still a finite probability for S=1 states\cite{alhassid2000b}. For increasing values of $r_s$, the probability for higher spin values increases and hereby the occurrence of spin pairs is less likely. It is expected that for typical values of $r_s\approx 1$ spin pairs are still likely to appear \cite{luescher2001, alhassid2000b, ullmo2008}. For our structure the electron gas interaction parameter can be estimated to be $r_s=1/(\sqrt{\pi\ensuremath{n_{\text{s}}}\xspace}\times a_\mathrm{B}^*)=0.8$, with the effective Bohr radius $a_\mathrm{B}^*=a_\mathrm{B}\times\epsilon_r\times m_0/m^*$, the Bohr radius $a_\mathrm{B}$, the relative dielectric constant $\epsilon_r$, the electron mass $m_0$ and the effective electron mass $m^*$. At low interaction strength the exchange energy can be calculated via the relation $\xi=\Delta\mathrm{E}\times r_s \times  \mathrm{ln}(1/r_s)/\sqrt{2\pi}$ valid for $r_s\ll1$ \cite{ihnbook1}. Here $r_s\lesssim1$, hence the obtained value $\xi\approx 13\ensuremath{\,\mu\mathrm{eV}}\xspace$ might be a lower bound for $\xi$ as for stronger interactions $r_s>1$ the exchange energy is expected to become a significant fraction of the single-particle level spacing of $\approx 180\,\mu\mathrm{eV}$\cite{ullmo2008}.

Unexpectedly, the exchange interaction seems negligible in the investigated regime of our device in contrast to previous results \cite{tarucha2000, fuhrer2003}. One possible reason for this could be shielding due to the top gates located only $34\,\mathrm{nm}$ above the 2DEG in this heterostructure. However, it is unlikely that this effect plays a major role as the top gates are not located exactly above the QD and the screening length of the 2DEG is estimated to be only around $32\,\mathrm{nm}$.\\

\subsection{Investigation of tunneling rates at zero magnetic field}
In Fig.\,\ref{figure2}(a) two other spin pairs (II/II' and III/III') are shown as a function of plunger gate voltage $V_\mathrm{PG}$ and perpendicular magnetic field $B$. The color code corresponds to the number of electrons entering and leaving the QD per second extracted from the QPC current measured as a function of time. From these time traces (see Fig.\,\ref{figure1}(b)), the tunneling rates $\ensuremath{\Gamma_\mathrm{in}=1/\langle\tau_\mathrm{in}\rangle}$, $\ensuremath{\Gamma_\mathrm{out}=1/\langle\tau_\mathrm{out}\rangle}$ can be extracted assuming a single-level model. For the peaks II/II' and III/III' the tunneling rates are shown in Fig.\,\ref{figure2}(b) at zero magnetic field. They are fitted by the single-level expressions $\ensuremath{\Gamma_\mathrm{in}=\tilde{\Gamma}_\mathrm{in}f(\mu)}$ and $\ensuremath{\Gamma_\mathrm{out}=\tilde{\Gamma}_\mathrm{out}(1-f(\mu))}$ using a maximum likelihood method, where the contribution of the individual data points is weighted by the inverse square of the statistical error. Here $\ensuremath{\tilde\Gamma_\mathrm{in}}$, $\ensuremath{\tilde\Gamma_\mathrm{out}}$ are constant fitting parameters and $f(\mu)$ is the Fermi function, where $\mu$ is the chemical potential. From the fits an upper limit for the electron temperature of $90\,\mathrm{mK}$ is extracted.

\subsection{Tunneling rates as a function of energy level detuning and magnetic field}
For further investigation, the tunneling rates are determined also as a function of perpendicular magnetic field $B$.
The results for pair III'/III (shown in Fig.\,\ref{figure3}(f),(g)) are presented in Fig.\,\ref{figure3}(a - e). In Fig.\,\ref{figure3}(a) the tunneling-out rate \ensuremath{\,\Gamma^{'}_{\mathrm{out}}}\xspace is plotted, where clear step-like transitions are visible (marked by arrows). A cut of Fig.\,\ref{figure3}(a) and (d) is presented in Fig.\,\ref{figure3}(c) to emphasize these step-like transitions in the tunneling-out rate of peak III'. The data corresponding to peak III are plotted in Fig.\,\ref{figure3}(b),\,(e), (g). The step-like features can be explained by a crossing of two energy levels\cite{tarucha2001}.  Additionally, four cuts at various magnetic fields are plotted in Fig.\,\ref{figure3}(h-k) and discussed later on in the paper. In the following section we focus on the investigation of the individual tunneling rates of single levels.

\section{Model to explain tunneling rates depending the on spin-configuration of the quantum dot}
\subsection{Model to explain tunneling rates considering a single energy level}
\label{subsecmodel}
Spin up and down electrons are filled alternately into a single spin-degenerate energy level of the QD in a model where the exchange interaction is neglected. The rates for tunneling into and out of the QD are expected to behave as depicted in Fig.\,\ref{figure4}(a,$\,$b) assuming zero bias voltage, equal electrochemical potential for $\uparrow$ and $\downarrow$ electrons in the QD ($\mu_{\uparrow}=\mu_{\downarrow}\equiv \mu$) and equal tunnel coupling for both spin species. For these assumptions the spin relaxation time has no influence on the resulting tunneling-in and tunneling-out rates. The red arrows mark options for a tunneling-in or -out event and the gray (white) filled circles refer to occupied (unoccupied) states. The first electron of the spin pair can either be a spin up or a spin down electron (see Fig.\,\ref{figure4}(a)). As soon as one electron tunneled into the QD, a second electron can only enter if the spin is opposite to the one of the first electron of the spin pair and if it has a larger energy due to the extra charging energy needed (see Fig.\,\ref{figure4}(b)). If the QD is occupied with the first electron of the spin pair, this electron has the chance to tunnel out. If however both the spin-up and the spin-down electron are filled into the QD either of them can tunnel out. Therefore the tunneling-in rate for the first electron of the spin pair is expected to be twice as large as the tunneling-out rate and vice versa for the second electron of the spin pair\cite{Gustavsson2009}. Here we used the additional assumption that the tunnel coupling is independent of any extrinsic parameter such as gate voltage and magnetic field.


\subsection{Comparison of model and measured tunneling rates at zero magnetic field}
Even though there is evidence that spin pairs are present in the investigated QD regime due to the parallel Coulomb peak evolution at finite magnetic field, the simple model presented in Fig.\,\ref{figure4}(a,$\,$b) only predicts the rates observed in the experiment approximately.

As an example we consider the peaks II/II' and III/III' in Fig.\,\ref{figure2}(b). The first electron of the spin pair is marked with a prime symbol while the second electron has a label without prime symbol. For peak III (\ensuremath{\,\Gamma_{\mathrm{in}}}\xspace,\ensuremath{\,\Gamma_{\mathrm{out}}}\xspace) and III' (\ensuremath{\,\Gamma\xspace '_{\mathrm{in}}}\xspace,\ensuremath{\,\Gamma\xspace '_{\mathrm{out}}}\xspace) we find: (\ensuremath{\,\tilde{\Gamma}\xspace '_{\mathrm{in}}}/\ensuremath{\,\tilde{\Gamma}\xspace '_{\mathrm{out}}}\xspace= $1.65\pm0.02$, \ensuremath{\,\tilde{\Gamma}_{\mathrm{out}}}\xspace/\ensuremath{\,\tilde{\Gamma}_{\mathrm{in}}}\xspace=$1.76\pm0.02$ ,\ensuremath{\,\tilde{\Gamma}\xspace '_{\mathrm{in}}}/\ensuremath{\,\tilde{\Gamma}_{\mathrm{in}}}\xspace= $1.41\pm0.03$,\ensuremath{\,\tilde{\Gamma}_{\mathrm{out}}}\xspace/\ensuremath{\,\tilde{\Gamma}\xspace '_{\mathrm{out}}}\xspace=$2.07\pm0.02$). Similarly for the other presented pair (II, II') we find ratios of $3.03\pm0.12$, $1.91\pm0.06$, $1.89\pm0.08$ and $3.07\pm0.09$. Investigating the tunneling rates of 12 pairs/triples at zero magnetic field we find that most of them do not behave precisely as predicted by the model, meaning that the ratio $\tilde{\Gamma}_\mathrm{in}/\tilde{\Gamma}_\mathrm{out}$ of the tunneling rates is often neither exactly 2 nor 1/2. \\


Possible reasons why the model does not hold at zero magnetic field are discussed in the following. 
The wave function overlap of an electron in the QD with the lead might depend on the spin with the consequence that $\ensuremath{\,\tilde{\Gamma}_{\mathrm{in}\,\uparrow}}\xspace=\ensuremath{\,\tilde{\Gamma}_{\mathrm{in}\,\downarrow}}\xspace$ is not necessarily true. However, the effect that electrons with a specific spin direction ($\uparrow$ or $\downarrow$) prefer to tunnel would only bring the ratios of tunneling rates closer to one. Hence, this argument does not qualify for explaining the measured tunneling rates. 
The capacitive cross-talk of the plunger gate to the source and drain barrier can have an influence on the tunneling rates as well. The barriers are expected to close slightly for more negative gate voltage values and hence we would expect $\ensuremath{\,\tilde{\Gamma}\xspace '_{\mathrm{in}}}/\ensuremath{\,\tilde{\Gamma}\xspace '_{\mathrm{out}}}\xspace\gtrsim2$, $\ensuremath{\,\tilde{\Gamma}_{\mathrm{out}}}\xspace/\ensuremath{\,\tilde{\Gamma}_{\mathrm{in}}}\xspace\lesssim2$, $\ensuremath{\,\tilde{\Gamma}\xspace '_{\mathrm{in}}}/\ensuremath{\,\tilde{\Gamma}_{\mathrm{in}}}\xspace\lesssim2$ and $\ensuremath{\,\tilde{\Gamma}_{\mathrm{out}}}\xspace/\ensuremath{\,\tilde{\Gamma}\xspace '_{\mathrm{out}}}\xspace\gtrsim2$. This model does not fully explain the experimental observations either. 
As there are at least $160-200$ electrons in the QD, multi-level transport due to orbital degeneracies might be another reason for the measured tunneling rates at zero magnetic field partly shown in Fig.\,\ref{figure2}.\\ 

\subsection{Comparison of model and measured tunneling rates at finite magnetic field}
The tunneling rates at small finite magnetic fields are analyzed using the data shown in Fig.\,\ref{figure3}. For the first electron of the spin pair the extracted ratios of the tunneling rates $\ensuremath{\tilde{\Gamma}\xspace '_\mathrm{in}/\tilde{\Gamma}\xspace '_\mathrm{out}}$ are $1.75 \pm 0.05$ (Fig.\,\ref{figure3}(i)) and $1.62 \pm 0.05$ (Fig.\,\ref{figure3}(j)), also marked in Fig.\,\ref{figure3}(a,d) by dash-dotted white lines. For the second electron the ratios of the tunneling rates $\ensuremath{\tilde{\Gamma}_\mathrm{out}/\tilde{\Gamma}_\mathrm{in}}$ are $1.21 \pm 0.03$ (Fig.\,\ref{figure3}(h)) and $2.30 \pm 0.07$ (Fig.\,\ref{figure3}(k)), also marked in Fig.\,\ref{figure3}(b,e) by dash-dotted white lines. Even though the measured values are around the expected ratio of $\ensuremath{\tilde{\Gamma}\xspace '_\mathrm{in}/\tilde{\Gamma}\xspace '_\mathrm{out}=\tilde{\Gamma}_\mathrm{out}/\tilde{\Gamma}_\mathrm{in}=2}$, significant deviations can be identified. We therefore conclude that the assumptions for the model described in Sec.\,\ref{subsecmodel} and Fig.\,\ref{figure4} are not entirely valid for our measurements. Specifically, the dependence of the tunneling rate on the magnetic field and on the energy is not included in the model. In addition we observe signs of resonances in the leads. A potential candidate for such a resonant state in the leads is marked by an orange arrow in Fig.\,\ref{figure3}(b/h).\\


\section{Conclusion}
The presented data reveals many consecutive spin pairs in a regime where at least 160-200 electrons populate the QD. Surprisingly, all investigated Coulomb peaks appear in pairs (or in triples in the case where one of the peaks is measured twice due to a charge rearrangement). One possible reason why exchange interaction is negligible for these measurements might be the shielding by the top gates located only $34\,\mathrm{nm}$ above the 2DEG in this heterostructure. Furthermore, the tunneling rates are investigated as a function of plunger gate voltage and magnetic field. A comparison of the data with a standard model reveals discrepancies between data and expectations. However, resonances in the leads and energy dependent tunnel barriers seem to be promising candidates to explain the deviation. 

\section{Acknowledgments}
The authors wish to thank Julien Basset, Simon Gustavsson, Andrea Hofmann, Bruno K\"ung, Ville Maisi, Clemens R\"ossler and Lars Steffen for their help. This research was supported by the Swiss National Science Foundation through the National Centre of Competence in Research Quantum Science and Technology (NCCR-QSIT).

\section*{References}

\begin{thebibliography}{99}
\bibitem{comment1}
The spin pairs are occasionally appearing as triples due to charge
  rearrangements.
\bibitem{comment3}
There is no clear odd-even effect in the spacing between conduction peaks present over the full range, however sometimes for 2-3 pairs one could argue that this 	  effect is visible. There are two possible reasons why this effect is not clearly seen:
On the one hand the single-particle level spacing in this many-electron dot is small compared to the charging energy, and on the other hand the effect may be masked by charge rearrangements.
\bibitem{comment4}
Parallel shifting peaks can indicate parallel spins, if the peak shifts are due to the Zeeman effect. In carbon nanotubes the Zeemann effect has a major impact on the magnetic field dependence, while for our device the Zeeman effect can be neglected in the magnetic field range investigated.
\bibitem{comment2}
For several such situations the correlation between triples of peaks and
  charge rearrangements was experimentally confirmed by measuring Coulomb
  diamonds where charge rearrangements are visible over the entire bias voltage
  range. The reasons why triples are observed are the following: A strongly coupled charge trap as present in this sample can shift the dot spectrum by almost one electron for each electron leaving or entering the charge trap. The plunger gate is not only coupled to the quantum dot but also to the charge trap. For example by applying a more negative voltage to the plunger gate, the charge trap is emptied at a certain point. To reach the same electrostatic configuration in the quantum dot that was present before this change of the charge trap, an even more negative voltage has to be applied to the plunger gate. Hence, a certain range of the electrostatic potential in the dot is measured twice leading to a triple.
\end{thebibliography}

\end{document}